\documentclass[journal,10pt,twocolumn]{IEEEtran}
\usepackage{amsmath,amsfonts}
\usepackage{algorithmic}
\usepackage{algorithm}
\usepackage{array}
\usepackage[caption=false,font=normalsize,labelfont=sf,textfont=sf]{subfig}
\usepackage{textcomp}
\usepackage{stfloats}
\usepackage{url}
\usepackage{verbatim}
\usepackage{graphicx}
\usepackage{cite}
\usepackage{booktabs}
\usepackage{multirow}
\usepackage{color}
\usepackage{tabularx}
\usepackage{hyperref}

\begin{document}

\title{From GDSII to Wafer: EDA Design Flow and Data Conversion for Wafer-Scale Superconducting Quantum Chip Fabrication}

\author{\IEEEauthorblockN{Ling Qiao\textsuperscript{1,2}, Fumin Luo\textsuperscript{1,2}, and Qinglang Guo\textsuperscript{1,2}*}\\[6pt]
\IEEEauthorblockA{\textsuperscript{1}Yangtze Delta Industrial Innovation Center of Quantum Science and Technology, Suzhou, China, 215100\\[2pt]
\textsuperscript{2}China Academy of Electronics and Information Technology, No.\ 11 Shuangyuan Road,\\
Shijingshan District, Beijing, China, 100041\\[2pt]
qiaoling@tgqs.net, 1763098000@qq.com\\[2pt]
*Corresponding author: gql1993@mail.ustc.edu.cn}}

\maketitle

\begin{abstract}
Superconducting quantum computing is advancing toward the thousand- and even million-qubit regime, making wafer-scale fabrication an essential pathway for achieving large-scale, cost-effective quantum processors. This manufacturing paradigm imposes new requirements on quantum-chip electronic design automation (Q-EDA): design tools must not only generate layouts (GDSII files) that satisfy quantum-circuit physical constraints but also ensure that the design data can be seamlessly converted into a complete set of manufacturing files executable by a wafer foundry, thereby enabling reliable translation from design intent to physical chip. This paper focuses on this critical data-conversion pipeline and presents a systematic treatment of the Q-EDA technology stack for wafer-scale fabrication. Starting from GDSII as the single authoritative data source, we analyze the key stages including process-design-kit (PDK)-based design rule checking (DRC), layout-versus-schematic (LVS) verification, design for manufacturability (DFM) optimization, wafer layout planning, and mask data preparation (MDP). We describe the concrete architecture of a Q-EDA system, present nine quantum-specific DRC rules together with their physical underpinnings and a multi-layer process stack model, and benchmark the manufacturing data-flow coverage of mainstream Q-EDA tools. Finally, we discuss the core challenges and future directions in this field.
\end{abstract}

\begin{IEEEkeywords}
Superconducting quantum chip, electronic design automation (EDA), wafer-scale fabrication, GDSII, process design kit (PDK), mask data preparation (MDP)
\end{IEEEkeywords}

\section{Introduction: Design-Data Challenges in the Era of Wafer-Scale Fabrication}

With the successive debut of superconducting quantum processors such as ``Origin Wukong'' (72 working qubits)~\cite{zhao2025eda} and ``Tianyan-504'' (504 qubits), quantum computing hardware is rapidly transitioning from laboratory prototypes to engineered systems~\cite{arute2019quantum}. To reach the million-qubit scale required for practical quantum computation, wafer-scale fabrication technology must be adopted, whereby multiple quantum chips---or a single ultra-large-scale chip---are manufactured in one pass on 300\,mm (12-inch) wafers, substantially improving fabrication throughput and reducing per-qubit cost~\cite{van2024advanced}. In 2024, IMEC successfully demonstrated the fabrication of superconducting transmon qubits on a 300\,mm CMOS production line, achieving relaxation and coherence times both exceeding 100\,$\mu$s, marking the transition of wafer-scale quantum chip fabrication from proof-of-concept to engineering practice~\cite{van2024advanced}.

However, wafer-scale fabrication extends the endpoint of chip design far beyond the conventional ``tape-out'' (delivery of a GDSII file) to a ``fab-out'' (generation of a complete set of manufacturing files capable of driving the production line). This poses unprecedented challenges for Q-EDA tools.

First, \textbf{data completeness} is required. A single GDSII file contains only the geometric information of each layer. Wafer fabrication demands a suite of derived files---layer-separated mask data, process-step control files, test-structure layouts, wafer dicing maps, and more~\cite{gdsii_format}---that collectively form the complete instruction set for lithography tools, etch systems, thin-film deposition equipment, and other semiconductor fabrication apparatus.

Second, \textbf{tight process binding} is essential. The design must be deeply coupled to the process capabilities of a specific fabrication line. Any design-rule violation or process mismatch can render an entire wafer defective, incurring substantial cost. Superconducting quantum chip fabrication involves thin-film deposition of niobium (Nb) or aluminum (Al), electron-beam lithography (EBL), and precision oxidation control of Josephson junctions (JJs)~\cite{gao2021practical}. These processes are highly sensitive to layout geometry. For instance, the critical current $I_c$ of a JJ is proportional to its physical area, and deviations in $I_c$ directly translate into qubit frequency shifts~\cite{zheng2023fabrication}.

Third, \textbf{manufacturability verification must be front-loaded}. Prior to data delivery, process-variation impacts on key qubit performance metrics (e.g., frequency and coherence time) must be predicted through simulation, and the design must be optimized accordingly~\cite{levenson2024review}. The mature DFM (Design for Manufacturability) methodology from the classical CMOS domain requires adaptation and extension to address the unique physics of superconducting quantum chips.

Therefore, the core mission of a modern Q-EDA tool is to build an end-to-end automated data pipeline anchored on GDSII, ensuring that quantum chip designs are accurately and efficiently converted into executable manufacturing instructions. This paper analyzes the key stages and technical underpinnings of this data flow, and presents the complete data-conversion chain from design to fabrication through the architecture of a Q-EDA system developed by the authors.

\section{Q-EDA System Architecture: A Data-Flow-Driven Back-End Engine}

To support the complete data flow from GDSII to manufacturing files, we designed a Q-EDA system architecture oriented toward wafer-scale fabrication. The architecture abandons concepts from traditional printed-circuit-board (PCB) EDA that are inapplicable to this domain (e.g., Gerber output, drill files) and focuses entirely on semiconductor microfabrication workflows~\cite{krylov2021review}.

As shown in Fig.~\ref{fig:system_arch}, the Q-EDA system architecture comprises four core layers, with data flowing top-down:

\begin{figure}[htbp]
\centerline{\includegraphics[width=0.95\columnwidth]{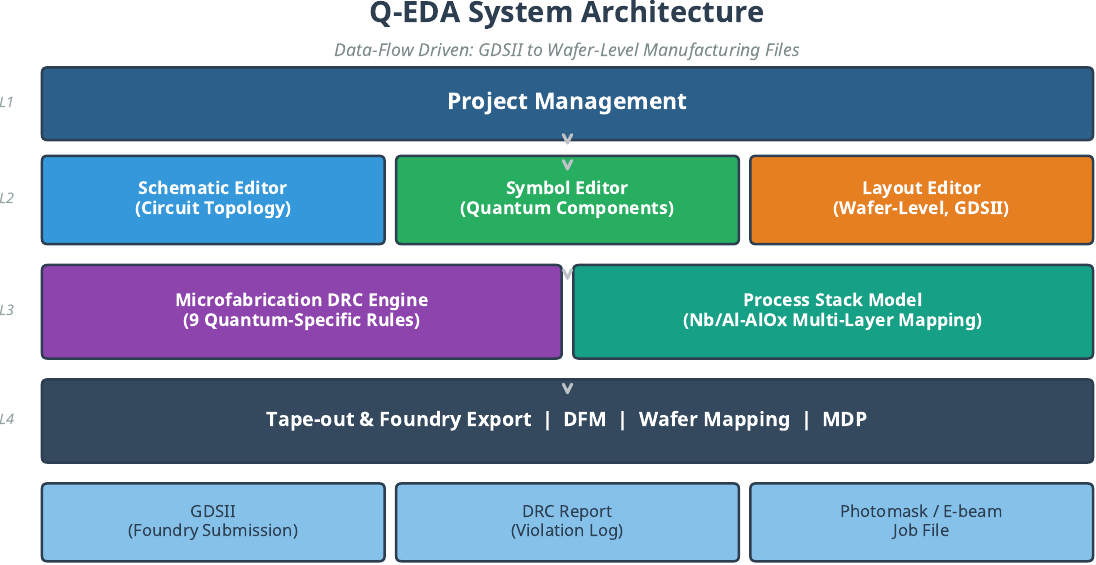}}
\caption{Q-EDA system architecture for wafer-scale quantum chip design. The layered architecture comprises: (1) Project Management; (2) Design Capture (schematic, symbol, and GDSII layout editors); (3) Verification (microfabrication DRC engine and process stack model); (4) Tape-out \& Foundry Export (manufacturing data preparation).}
\label{fig:system_arch}
\end{figure}

\textbf{Layer~1: Project Management.} This layer handles global design-parameter control and version management, ensuring data synchronization between schematics and physical layouts. In a wafer-scale fabrication context, project management must additionally coordinate the design parameters of multiple die variants, manage design configurations across different process corners, and maintain full traceability of the design history.

\textbf{Layer~2: Design Capture.} This layer contains the schematic editor, symbol editor, and layout editor. The layout editor serves as the core physical-design environment, operating directly on the GDSII stream format with nanometer-level precision~\cite{gdsii_format} and converting parameterized models of quantum components into precise geometric polygons. Q-EDA incorporates a rich library of quantum components---including transmon qubits, Xmon qubits, coplanar-waveguide (CPW) resonators, couplers, and Purcell filters~\cite{kjaergaard2020superconducting}---each implemented as a parameterized cell (P-cell). Users need only adjust key physical parameters (e.g., JJ area, CPW impedance, resonant frequency) for the layout to be generated automatically~\cite{qiskitmetal2021}.

\textbf{Layer~3: Verification \& Process Mapping.} This layer bridges design and fabrication. It contains the microfabrication DRC engine and the process stack model. The former executes quantum-specific design-rule checks~\cite{van2019layout}; the latter maps logical GDSII layers to physical fabrication steps (e.g., Nb base electrode, AlO$_x$ barrier layer)~\cite{tolpygo2016advanced}. Together, they ensure that the design data is fully compliant before entering the manufacturing-preparation stage.

\textbf{Layer~4: Tape-out \& Foundry Export.} This layer performs the final data conversion, including wafer layout planning (wafer mapping), mask data preparation (MDP), and ultimately outputs the foundry-required deliverables: the GDSII submission package, DRC violation reports, and job files for lithography or electron-beam direct-write tools.

In the following sections, we trace this data flow and analyze each key stage from GDSII input to final manufacturing-file generation.

\section{GDSII Input and PDK-Based Physical Verification}

A complete Q-EDA data flow supporting wafer-scale fabrication begins with the quantum chip layout (GDSII) and ends with manufacturing files that drive lithography tools, etch systems, and related equipment. This section focuses on the starting point and the first gate of the data flow: GDSII input and PDK-based physical verification.

\subsection{Starting Point: The GDSII Layout File}
GDSII (Graphic Data System~II) is a binary database file format that has become the de facto industry standard for describing integrated-circuit physical layouts~\cite{gdsii_format}. In superconducting quantum chip design, the GDSII file defines the precise geometry, hierarchical relationships, and layer attributes of all structures---transmon capacitor pads, Josephson junction regions, CPW resonators, control lines, and readout lines. It serves as the single authoritative source for all subsequent manufacturing data processing.

GDSII files organize data hierarchically: a top cell contains cell references to sub-cells, which can be instantiated multiple times to efficiently represent repetitive structures (e.g., repeating unit cells in a qubit array). Each cell contains various geometric primitives---boundaries (polygons), paths, and text annotations---each associated with a specific layer and datatype to distinguish different fabrication process steps.

The Q-EDA GDSII parsing engine is built on a C++ geometry kernel (gdstk), enabling high-speed read/write of large-scale layout files. In prototype validation, the DRC engine completed 76 geometric rule checks on a test chip containing 28 parameterized components in only 0.23\,ms, demonstrating sub-millisecond response times (see Section~VII for detailed results)~\cite{gdsfactory2023}.

Compared with existing approaches, the Q-EDA data-flow architecture presented in this paper offers the following core advantages: (i)~it adopts a \emph{single-source-of-truth} principle, using GDSII as the sole authoritative input, thereby avoiding information loss and consistency risks introduced by data-format conversions across multi-tool chains; (ii)~PDK rules are encoded in structured JSON, supporting hot-swappable process rules for different foundries without modifying the engine code; (iii)~physical verification (DRC/LVS/ERC), process stack mapping, DFM optimization, and mask data preparation are unified into a single automated pipeline, eliminating the efficiency bottlenecks and human-error risks inherent in the manual handoffs of traditional workflows; (iv)~the underlying C++ geometry engine, combined with spatial indexing (R-tree) optimization, guarantees linear ($O(n)$) time complexity for DRC checks, providing a scalable technical foundation for future large-scale quantum chip layout verification.

\subsection{Process Design Kit (PDK): The Contract Between Design and Fabrication}
Before entering the manufacturing data-conversion stage, the GDSII must pass a series of rigorous physical verification steps to confirm compliance with the fabrication process's design rules and consistency with the original circuit design intent. This process relies heavily on the process design kit (PDK)~\cite{kohli2023white}.

A PDK is a standardized toolkit provided by the wafer foundry and serves as the core ``contract'' between chip designers and the fabrication process. A complete PDK typically includes the following elements:

\textbf{Process parameter files:} These define the material properties (e.g., resistivity, dielectric constant), thickness ranges, and etch biases of each thin-film layer. For superconducting quantum chips, these parameters directly affect CPW characteristic impedance, resonant frequency, and qubit coherence time~\cite{goppl2008coplanar}.

\textbf{Device model library:} This provides physical and equivalent-circuit models for fundamental devices such as Josephson junctions, capacitors, and inductors, used for circuit simulation and parameter extraction.

\textbf{Design rule files:} These encode all geometric constraints in a machine-readable format (e.g., Calibre DRC rule files or custom scripts), serving as the basis for DRC engine execution.

\textbf{Parameterized cell library (P-cell):} This provides pre-verified, parameterizable standard component layouts that designers can directly invoke and customize, avoiding the need to draw layouts from scratch.

CMC Microsystems, in its superconducting PDK white paper~\cite{kohli2023white}, systematically defined a PDK framework for superconducting quantum circuits, covering the full spectrum from material parameters to design rules. This provides an important reference for industry standardization.

\subsection{Key Verification Steps}
Based on the PDK, physical verification comprises three core steps:

\textbf{Design Rule Checking (DRC):} This automatically verifies that all geometric relationships in the layout---widths, spacings, enclosures, overlaps---satisfy the minimum feature-size requirements of the process line. DRC is the most fundamental and critical verification step; any violation indicates that the design cannot be correctly fabricated~\cite{fourie2020eda}.

\textbf{Layout Versus Schematic (LVS):} This compares the electrical connectivity netlist extracted from the physical layout against the original schematic, ensuring that the two are fully consistent in terms of Josephson junction count and connectivity~\cite{van2019layout}. LVS extracts parasitic parameters and connectivity from the layout to generate a ``layout netlist,'' which is then compared node-by-node and device-by-device against the schematic netlist.

\textbf{Electrical Rule Checking (ERC):} This checks for potential electrical issues such as short circuits, open circuits, floating nodes, and antenna effects. In superconducting quantum chips, ERC must additionally verify ground-plane continuity, because a discontinuous ground plane introduces parasitic slotline modes that severely degrade microwave signal transmission quality~\cite{huang2021microwave}.

\subsection{Quantum-Specific DRC Rule Engine for Manufacturing}
Unlike traditional CMOS processes, which are primarily constrained by lithographic resolution, the DRC rules for superconducting quantum chips are deeply rooted in microwave engineering and quantum coherence physics~\cite{levenson2024review}. The Q-EDA microfabrication DRC engine implements nine quantum-specific design rules, as listed in Table~\ref{tab:drc_rules}.

\begin{table}[htbp]
\caption{Quantum-Specific DRC Rule Set Implemented in Q-EDA}
\label{tab:drc_rules}
\begin{center}
\footnotesize
\setlength{\tabcolsep}{3pt}
\begin{tabular}{@{}cp{2.6cm}cc@{}}
\toprule
\textbf{ID} & \textbf{Rule Description} & \textbf{Threshold} & \textbf{Physical Origin} \\ \midrule
R1 & CPW gap min.\ width & $\ge 3$\,$\mu$m & TLS loss control \\
R2 & CPW conductor width & $\ge 5$\,$\mu$m & Impedance matching \\
R3 & JJ overlap tolerance & $\pm 50$\,nm & $I_c$ control \\
R4 & JJ lead min.\ width & $\ge 100$\,nm & E-beam resolution \\
R5 & Airbridge span range & 50--100\,$\mu$m & Mech.\ stability \\
R6 & Airbridge pad size & $\ge 10{\times}10$\,$\mu$m & Adhesion \\
R7 & Chip-edge clearance & $\ge 200$\,$\mu$m & Dicing stress \\
R8 & Metal-to-metal spacing & $\ge 2$\,$\mu$m & Short-circuit prev. \\
R9 & Ground-plane continuity & No gap $> 50$\,$\mu$m & Parasitic mode sup. \\ \bottomrule
\end{tabular}
\end{center}
\end{table}

The physical background and engineering significance of each rule are analyzed below.

\textbf{CPW gap minimum width (R1):} The coplanar waveguide (CPW) is the fundamental structure for transmitting microwave signals in superconducting quantum chips. The substrate--metal interface exposed in the gap region is the primary source of two-level system (TLS) defects~\cite{place2021new}. TLS defects absorb microwave photon energy, causing energy relaxation of the qubit (reduced $T_1$). A narrower gap increases the electric-field intensity at the interface, raises the electric-field participation ratio, and exacerbates TLS-induced dielectric loss. Consequently, the CPW gap must exceed 3\,$\mu$m to keep the participation ratio within acceptable bounds.

\textbf{CPW center conductor width (R2):} The characteristic impedance of a CPW is determined by the ratio of center conductor width $w$ to gap width $s$, typically designed to be 50\,$\Omega$ for impedance matching with external microwave measurement systems~\cite{goppl2008coplanar}. An excessively narrow center conductor not only increases nonlinear effects introduced by kinetic inductance but also raises the current density, reducing the superconducting critical-current margin.

\textbf{JJ overlap alignment tolerance (R3):} The transmon qubit frequency $f_{01} \approx \sqrt{8 E_J E_C}/h$ is highly sensitive to the Josephson energy $E_J$, where $E_J = I_c \Phi_0 / (2\pi)$ is proportional to the critical current $I_c$~\cite{kjaergaard2020superconducting}. Since $I_c$ is proportional to the physical area of the JJ, the overlap area between the upper and lower electrodes in an overlap-process JJ is determined by lithographic alignment accuracy. The typical alignment error for electron-beam lithography (EBL) is $\pm 50$\,nm~\cite{gao2021practical}. The DRC engine must verify that the designed overlap region provides sufficient margin to absorb this alignment error, ensuring that $I_c$ deviation remains within acceptable bounds (typically $\Delta I_c / I_c < 5\%$)~\cite{zheng2023fabrication}.

\textbf{JJ lead minimum width (R4):} The width of leads connecting a JJ to the CPW or capacitor pad is limited by the resolution of electron-beam lithography (typically 50--100\,nm). Excessively narrow leads increase resistance (even in the superconducting state, nanoscale constrictions can introduce additional kinetic inductance) and raise the probability of fabrication defects.

\textbf{Airbridge span constraint (R5):} Airbridges connect the ground planes on both sides of a CPW to suppress parasitic slotline modes~\cite{chen2014fabrication}. They are typically fabricated using a sacrificial-layer process: a photoresist is first deposited as a support, metal is then evaporated to form the bridge, and finally the photoresist is dissolved to release the suspended structure. A span exceeding 100\,$\mu$m risks collapse of the suspended film during sacrificial-layer release due to gravity or surface tension; a span that is too short fails to bridge the CPW center conductor, negating its parasitic-mode suppression function.

\textbf{Airbridge pad size (R6):} The anchor pads at both ends of an airbridge must be sufficiently large ($\ge 10 \times 10$\,$\mu$m) to provide adequate metal--metal contact area and adhesion, preventing detachment during subsequent process steps (e.g., ultrasonic cleaning, cryogenic thermal cycling).

\textbf{Chip-edge clearance (R7):} During wafer dicing, diamond saw blades or lasers generate mechanical stress and micro-cracks near the scribe lanes. All functional features must be at least 200\,$\mu$m from the chip edge to prevent thin-film delamination or structural damage caused by stress propagation.

\textbf{Metal-to-metal spacing (R8):} Same-layer metal features must maintain a minimum spacing of 2\,$\mu$m to prevent short circuits caused by incomplete etching or metal residues. This rule accounts for both the resolution limit of optical lithography and the isotropic bias of the etch process.

\textbf{Ground-plane continuity (R9):} The ground plane of a superconducting quantum chip must remain electrically continuous. Any gap exceeding 50\,$\mu$m can form a parasitic resonant cavity whose resonant frequency may overlap with the operating frequencies of qubits or readout resonators, causing severe crosstalk and decoherence~\cite{huang2021microwave}.

\subsection{Benchmarking Against the CMC Superconducting PDK White Paper}
The superconducting PDK white paper published by CMC Microsystems~\cite{kohli2023white} defines a standardized design-rule framework for both academic and industrial communities. Table~\ref{tab:pdk_benchmark} benchmarks the Q-EDA DRC rules against the CMC PDK rules.

\begin{table}[htbp]
\caption{Benchmarking of Q-EDA DRC Rules Against the CMC PDK White Paper}
\label{tab:pdk_benchmark}
\begin{center}
\footnotesize
\begin{tabular}{@{}lcc@{}}
\toprule
\textbf{Rule Category} & \textbf{CMC PDK} & \textbf{Q-EDA} \\ \midrule
CPW minimum gap & $\ge 2$\,$\mu$m & $\ge 3$\,$\mu$m \\
CPW minimum linewidth & $\ge 4$\,$\mu$m & $\ge 5$\,$\mu$m \\
JJ alignment tolerance & $\pm 100$\,nm & $\pm 50$\,nm \\
Metal minimum spacing & $\ge 2$\,$\mu$m & $\ge 2$\,$\mu$m \\
Edge clearance & $\ge 100$\,$\mu$m & $\ge 200$\,$\mu$m \\
Airbridge rules & Not defined & Fully defined \\
Ground-plane continuity & Recommended & Enforced \\ \bottomrule
\end{tabular}
\end{center}
\end{table}

The benchmarking reveals that Q-EDA adopts stricter thresholds for several rules (e.g., CPW gap of 3\,$\mu$m vs.\ 2\,$\mu$m for CMC), reflecting more stringent requirements for TLS loss control. Additionally, Q-EDA supplements the airbridge rules not yet covered by the CMC PDK and elevates ground-plane continuity from ``recommended'' to ``enforced,'' accommodating the design requirements of high-coherence qubits.

\section{Multi-Layer Process Stack Modeling and Process Mapping}

After physical verification, the GDSII data must be mapped to specific fabrication process steps. The physical realization of superconducting quantum chips requires complex, multi-layer microfabrication processes~\cite{tolpygo2016advanced}. To ensure manufacturability, Q-EDA integrates a comprehensive multi-layer process stack model that maps logical GDSII layers to physical fabrication steps.

\subsection{Process Stack Architecture}
A typical superconducting quantum chip process stack (Table~\ref{tab:process_stack}) comprises multiple critical layers, each with a specific physical function and stringent fabrication tolerances~\cite{gao2021practical}.

\begin{table*}[htbp]
\caption{Superconducting Quantum Chip Process Stack Parameters Modeled in Q-EDA}
\label{tab:process_stack}
\begin{center}
\begin{tabular}{@{}llccl@{}}
\toprule
\textbf{Layer} & \textbf{Material} & \textbf{Typical Thickness} & \textbf{Lithography} & \textbf{Physical Function} \\ \midrule
Substrate & c-plane sapphire (Al$_2$O$_3$) & 430\,$\mu$m & N/A & Mechanical support; low-loss dielectric ($\epsilon_r \approx 10$) \\
Base metal (M1) & Niobium (Nb) & 100--200\,nm & Optical (i-line) & Ground plane, CPW conductors, qubit capacitor pads \\
JJ bottom electrode & Nb or Al & 100--200\,nm & E-beam & Bottom electrode of the Josephson junction \\
Tunnel barrier & Aluminum oxide (AlO$_x$) & 1--2\,nm & N/A (oxidation) & Nonlinear inductance barrier defining $I_c$ \\
JJ top electrode & Nb or Al & 50--100\,nm & E-beam & Top electrode completing the tunnel junction \\
Wiring layer (M2) & Niobium (Nb) & 200--300\,nm & Optical & Signal routing, airbridge anchor points \\
Airbridge layer & Aluminum (Al) & 150--300\,nm & Optical & Suspended ground-plane stitching structure \\ \bottomrule
\end{tabular}
\end{center}
\end{table*}

\textbf{Substrate selection:} The foundation is typically a c-plane sapphire (Al$_2$O$_3$) or high-resistivity silicon (HR-Si) substrate. Sapphire is widely adopted owing to its extremely low microwave dielectric loss ($\tan\delta < 10^{-6}$) at millikelvin temperatures~\cite{place2021new}. High-resistivity silicon substrates are favored in industrial-scale fabrication for their compatibility with CMOS processes~\cite{van2024advanced}. The substrate choice directly influences the effective dielectric constant and microwave propagation characteristics of CPWs.

\textbf{Base metal layer (M1):} Typically a 100--200\,nm Nb film deposited by magnetron sputtering and patterned using optical lithography and reactive-ion etching (RIE)~\cite{tolpygo2016advanced}. This layer forms the large-scale features: ground plane, CPW center conductors, and transmon capacitor pads. The superconducting transition temperature of Nb ($T_c \approx 9.2$\,K) is well above the typical qubit operating temperature ($\sim$20\,mK), providing a robust superconducting margin.

\textbf{Josephson junction trilayer:} The core nonlinear element of a qubit is the Josephson junction, typically fabricated using a Nb/Al-AlO$_x$/Nb trilayer process~\cite{tolpygo2016advanced} or an Al/AlO$_x$/Al double-angle evaporation (Dolan bridge or Manhattan) process~\cite{gao2021practical}. The critical parameter is the AlO$_x$ tunnel barrier, only 1--2\,nm thick, whose thickness and uniformity directly determine the magnitude and consistency of $I_c$~\cite{zheng2023fabrication}. In industrial 300\,mm processes, IMEC has adopted a fully optical-lithography overlap JJ process, avoiding the low-throughput bottleneck of electron-beam lithography~\cite{van2024advanced}.

\textbf{Wiring layer and airbridges:} The wiring layer (M2) provides additional signal-routing capability and interlayer interconnects. The airbridge layer is fabricated using a sacrificial-layer process to form suspended metal bridge structures that stitch the ground planes on both sides of CPWs~\cite{chen2014fabrication}.

\subsection{Interlayer Alignment and Registration Budget}
Multi-layer processes require precise alignment between successive lithographic steps. Alignment accuracy directly determines the reliability of interlayer connections and the consistency of device parameters.

Optical lithography typically achieves an alignment accuracy of $\pm 0.5$\,$\mu$m, while electron-beam lithography can reach $\pm 50$\,nm~\cite{gao2021practical}. The Q-EDA process model automatically verifies that all interlayer connections (e.g., vias connecting M1 and M2, or JJ electrode-to-M1 connections) have sufficient overlap margin to maintain electrical continuity under worst-case alignment errors.

Specifically, the registration budget is calculated as follows: for any connection between two layers, the minimum overlap $O_{\min}$ must satisfy $O_{\min} \ge O_{\text{design}} - \sigma_{\text{align}}$, where $O_{\text{design}}$ is the nominal designed overlap and $\sigma_{\text{align}}$ is the alignment error between the two layers (taken at the $3\sigma$ level). Q-EDA automatically encodes this constraint as an interlayer overlap rule within the DRC checks.

\subsection{GDSII Layer-to-Process-Step Mapping}
The Q-EDA process stack model maintains a mapping table from logical GDSII layer numbers to physical fabrication steps. For example, GDSII Layer~1 may map to ``Nb base-metal sputtering + lithography + RIE etching,'' and Layer~5 to ``Al evaporation + oxidation + Al evaporation (JJ trilayer).'' This mapping ensures that operations performed by layout designers at the logical level are accurately translated into process instructions for fabrication engineers, forming the foundation for seamless design--fabrication integration.

\section{Design for Manufacturability, Wafer Planning, and Mask Data Preparation}

After physical verification and process mapping, the design data enters the final preparation stage for wafer-scale fabrication. The core objective of this stage is to convert the ideal geometry of a single chip into wafer-level manufacturing files that can tolerate process variations and drive lithographic equipment.

\subsection{Design for Manufacturability (DFM) and Process Co-optimization}
Passing DRC/LVS only guarantees that a design is ``manufacturable'' but does not guarantee ``high yield.'' DFM aims to analyze and optimize the design to tolerate inherent process variations. For superconducting quantum chips, DFM faces unique challenges:

\textbf{Process variation modeling:} Statistical models must be established that relate fabrication fluctuations in critical dimensions (e.g., JJ area, AlO$_x$ barrier thickness, CPW linewidth) to electrical parameters such as qubit frequency, coupling strength, and coherence time~\cite{levenson2024review}. For example, IMEC's 300\,mm wafer-scale data show that the across-wafer coefficient of variation (CV) of JJ resistance can be controlled to within 3\%~\cite{van2024advanced}, yet this still corresponds to qubit frequency shifts of tens of MHz---non-negligible in multi-qubit systems requiring precise frequency allocation.

\textbf{Hotspot detection and repair:} This involves identifying layout regions that are sensitive to process variations and prone to defects (e.g., metal-residue risk in densely routed areas, bridging risk at extremely narrow gaps), and optimizing them by adjusting feature spacing or adding redundant structures.

\textbf{Optical proximity correction (OPC):} At nanometer scales, optical diffraction causes the actual transferred pattern to deviate from the mask pattern (e.g., corner rounding, line-end shortening). OPC pre-compensates the mask pattern with inverse distortions (e.g., adding serif features, adjusting hammerhead corners) to ensure that the pattern projected onto the wafer matches the design intent. For superconducting quantum chips, OPC is primarily applied to optical lithography layers (e.g., the M1 base-metal layer), while electron-beam lithography layers have relatively lower OPC requirements owing to their high resolution.

\subsection{Wafer Layout Planning (Wafer Mapping)}
For wafer-scale fabrication, the single-chip GDSII data must be ``mapped'' onto an entire wafer. Taking IMEC's process for fabricating superconducting qubits on 300\,mm wafers as an example~\cite{van2024advanced}, wafer layout planning involves the following key steps:

\textbf{Step \& Repeat:} Based on wafer size (e.g., 300\,mm diameter) and chip size (e.g., 24\,mm $\times$ 28\,mm), the optimal tiling layout is computed to maximize the number of chip copies within the usable wafer area. IMEC's 300\,mm process places 75 dies on a single wafer~\cite{van2024advanced}. The Q-EDA wafer mapping module automatically performs this computation and generates a wafer-level GDSII file containing all chip copies.

\textbf{Scribe lane and test-structure insertion:} Alignment marks, process control monitors (PCMs), and test devices (e.g., JJ test arrays, resistance test structures) are placed in the scribe lanes between chips~\cite{van2024advanced}. These structures are used for real-time alignment during fabrication, process-parameter monitoring, and post-fabrication chip testing. IMEC includes 20 sub-dies within each die, some dedicated to across-wafer JJ resistance and qubit parameter statistics~\cite{van2024advanced}.

\textbf{Wafer edge exclusion zone:} Process uniformity near the wafer edge is degraded by non-uniform resist coating, temperature gradients, and other factors. Wafer mapping must define an edge exclusion zone---typically a 3--5\,mm annular region from the wafer edge---within which no functional chips are placed.

\subsection{Mask Data Preparation (MDP)}
Mask data preparation (MDP) is the core step that converts verified layout data into instructions directly usable by lithographic tools, and represents the final stage of design-to-manufacturing data conversion.

\textbf{Data format conversion and fracturing:} Complex GDSII polygons are decomposed (fractured) into simple shapes (rectangles and trapezoids) that mask-writing tools (e.g., electron-beam direct-write systems such as MEBES or laser writers) can process, and converted into proprietary data formats (e.g., MEBES or OASIS.MASK). The quality of the fracturing algorithm directly affects mask write time and pattern fidelity.

\textbf{Reticle generation:} The layout data for each layer of the wafer-level design is converted into corresponding photomask (reticle) data. Each process step corresponds to one reticle. For step-and-repeat optical lithography, the reticle need only contain the pattern for a single exposure field; the lithography tool replicates the pattern across the wafer by stepping the wafer stage.

\textbf{Job deck generation:} A set of control files is generated to drive mask-fabrication equipment and lithographic tools. These files define the precise pattern of each reticle, exposure dose, focus offset, alignment strategy, and stepping parameters. At this point, the design data has been fully converted into ``manufacturing files'' ready for delivery to the wafer foundry's production line.

\section{Q-EDA Tool Practices Supporting Manufacturing File Generation}

Both domestic and international Q-EDA tools are evolving toward support for the complete manufacturing data flow. Table~\ref{tab:tool_comparison} compares the coverage of mainstream tools across the data-flow stages.

\begin{table}[htbp]
\caption{Comparison of Mainstream Q-EDA Tools Across Manufacturing Data-Flow Stages}
\label{tab:tool_comparison}
\begin{center}
\scriptsize
\begin{tabular}{@{}p{1.2cm}p{1.4cm}p{1.4cm}p{1.4cm}@{}}
\toprule
\textbf{Stage} & \textbf{Origin Pilot} & \textbf{EDA-Q} & \textbf{Qiskit Metal} \\ \midrule
GDSII generation & Parameterized P-cell & Open-source component library & Python-driven \\
PDK verification & Built-in DRC/LVS & External dependency & Calibre dependency \\
Process mapping & 3D process simulation & Basic support & Basic support \\
DFM/Wafer planning & Wafer planning & Planned & Not supported \\
MDP export & One-click export & Planned & Not supported \\ \bottomrule
\end{tabular}
\end{center}
\end{table}

\textbf{Origin Pilot:} As a representative domestic industrial software~\cite{zhao2025eda}, Origin Pilot has strengthened its design-to-fabrication coordination capabilities. It explicitly includes wafer planning and layout partitioning modules and can leverage a 3D process simulation module to assist manufacturability assessment, establishing a preliminary data chain from design to manufacturing preparation. Its PDK framework supports user-defined process parameters and design rules, providing flexibility for interfacing with different foundries.

\textbf{EDA-Q:} As an open-source Q-EDA tool~\cite{zhao2025eda}, EDA-Q targets the full flow from topology design to simulation optimization. Its open-source nature facilitates community-driven development of foundry-specific interfaces and manufacturing data-output modules. Support for DFM and MDP stages is currently under active development.

\textbf{Qiskit Metal:} IBM's open-source framework~\cite{qiskitmetal2021} focuses primarily on rapid front-end design and simulation prototyping. For mature manufacturing file generation and deep integration with foundry PDKs, it typically requires coupling with traditional EDA vendor tool chains (e.g., Cadence, Siemens EDA) for physical verification and MDP.

\textbf{KQCircuits:} Developed by IQM Finland, this open-source KLayout Python library~\cite{kqcircuits2022} provides a rich parameterized component library for superconducting quantum circuits. Built on the KLayout platform, it natively supports GDSII read/write and basic DRC functionality, but relies on external tools for back-end manufacturing data-flow stages such as wafer planning and MDP.

\textbf{GDSFactory:} A general-purpose Python GDSII design library~\cite{gdsfactory2023} offering flexible parameterized layout generation. Although not specifically designed for quantum chips, its powerful scripting capabilities have led to widespread adoption in the quantum chip design community. GDSFactory can integrate with external DRC tools and simulators but does not itself provide complete manufacturing data-flow support.

Overall, the current Q-EDA tool ecosystem exhibits a ``strong front-end, weak back-end'' characteristic: most tools have developed adequate capabilities for GDSII layout generation and basic DRC, but significant gaps remain in DFM optimization, wafer-level layout planning, and MDP---the manufacturing-oriented back-end stages. Bridging this gap is key to the industrialization of Q-EDA tools.

\section{Prototype System and End-to-End Data-Flow Validation}

To validate the GDSII-to-manufacturing-file data-flow architecture proposed in this paper, we developed a Q-EDA desktop prototype system. The prototype is built on Python and PyQt6, with the underlying geometry engine implemented as a C++ extension (gdstk) to ensure processing efficiency.

\subsection{End-to-End Data Flow}
\begin{figure}[htbp]
\centerline{\includegraphics[width=0.95\columnwidth]{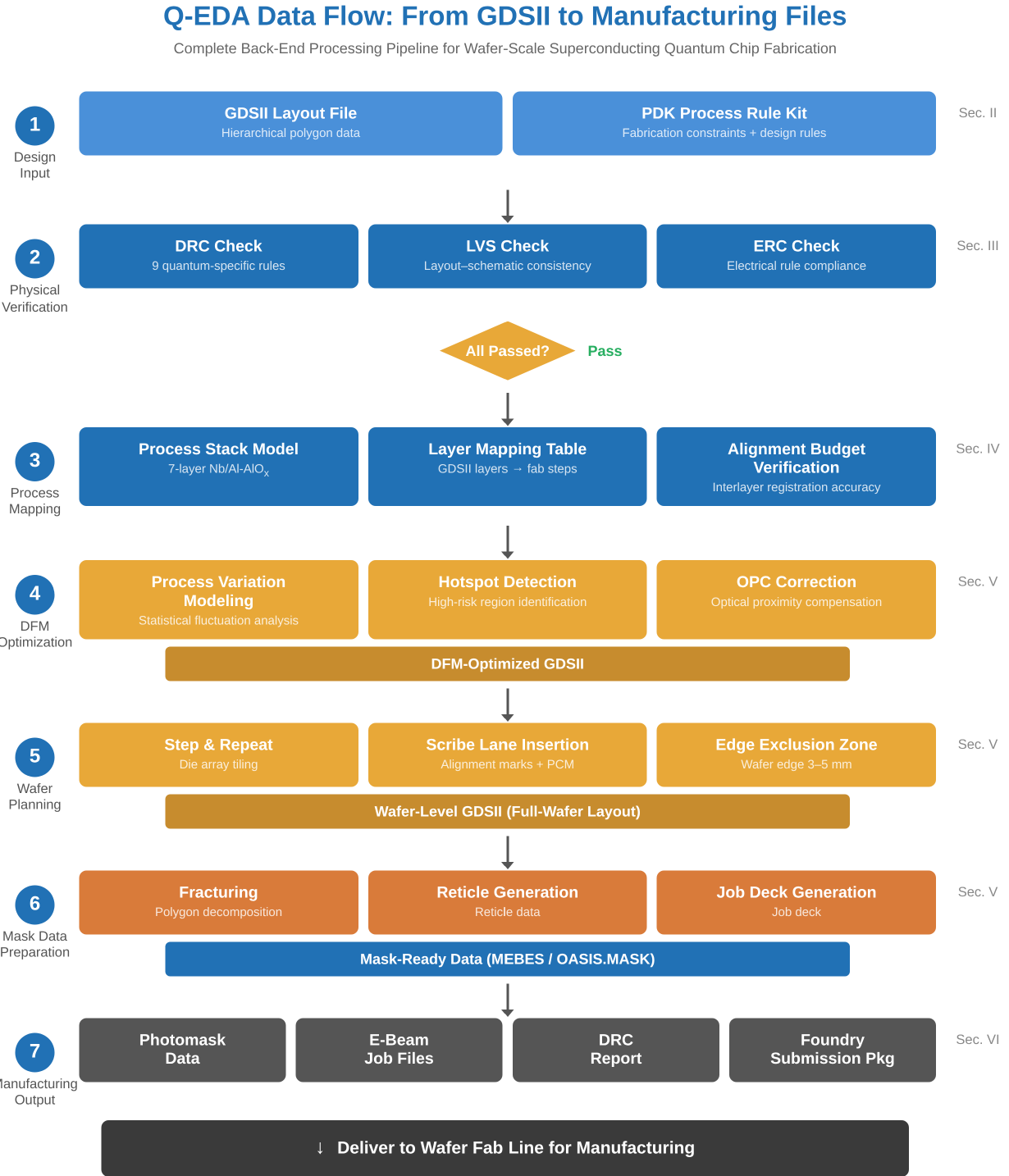}}%
\vspace{-2pt}
\caption{End-to-end data flow implemented by the Q-EDA prototype system, showing the seven complete stages from GDSII design input to final manufacturing file output.}
\label{fig:workflow}
\end{figure}

The prototype system fully implements the seven-stage data flow shown in Fig.~\ref{fig:workflow}.

In the prototype system, users can perform the following end-to-end operations through a graphical user interface (GUI):
\begin{enumerate}
\item \textbf{Parameterized design input:} Generate chip layouts containing Xmon qubits, CPW resonators, and airbridges via the component library panel.
\item \textbf{One-click physical verification:} Invoke the built-in DRC engine to execute nine quantum-specific rule checks on the generated GDSII layout, with real-time violation reporting and highlighting in the console.
\item \textbf{Process stack mapping:} Map logical layers to the seven-layer Nb/Al-AlO$_x$ physical fabrication steps through the process stack viewer, and compute interlayer alignment budgets.
\item \textbf{Tape-out readiness check and export:} Execute seven tape-out compliance checks---including nanometer-grid alignment, hierarchy flattening, and bounding-box integrity---and generate the final foundry submission package.
\end{enumerate}

\subsection{Prototype Validation Results}
We used the prototype system to generate a four-qubit diamond-topology test chip containing 28 parameterized components. The experimental results demonstrate that the prototype system can smoothly execute the complete data flow:
\begin{itemize}
\item \textbf{DRC verification:} 76 geometric rule checks were executed across 28 components in only 0.23\,ms, all passing with zero violations.
\item \textbf{Tape-out compliance:} All seven tape-out checks passed, and a complete manufacturing submission package---including the GDSII layout, DRC log, and process mapping table---was successfully generated.
\item \textbf{Processing efficiency:} Owing to the underlying C++ geometry engine and R-tree spatial indexing optimization, even in a stress test scaled to 50 qubits (1,016 geometric primitives), the full-chip DRC check completed in only 2.94\,ms, exhibiting strict linear ($O(n)$) scaling.
\end{itemize}

The successful operation of this prototype system validates the engineering feasibility of the proposed data-flow architecture and provides a verification platform for subsequent integration of more sophisticated DFM optimization and wafer-level MDP algorithms.

\section{Core Challenges and Future Outlook}

Although the data-flow pathway from GDSII to manufacturing files is becoming increasingly clear, achieving efficient and highly reliable wafer-scale fabrication still faces the following core challenges:

\textbf{Maturity of quantum-specific PDKs:} Compared with the mature PDK ecosystem accumulated over decades for traditional CMOS processes, superconducting quantum chip PDKs remain in their infancy. The lack of unified industry standards and production-validated kits leads to poor interoperability between different design teams and fabrication lines. The CMC Microsystems white paper~\cite{kohli2023white} represents an important step toward standardization, but a considerable gap remains before a CMOS-like complete PDK ecosystem can be established.

\textbf{Efficiency of large-scale data processing:} The layout data of a million-qubit chip may contain millions of geometric primitives. Performing full-wafer DRC, LVS, and OPC processing on such data poses severe demands on algorithmic efficiency and computational resources. Traditional serial processing approaches will be unable to meet engineering timelines, necessitating the development of parallelized algorithms and distributed computing frameworks tailored to quantum chip characteristics.

\textbf{Accurate modeling of process variations:} Superconducting quantum chips are far more sensitive to process variations than classical CMOS devices. Establishing accurate statistical models that correlate across-wafer and wafer-to-wafer process fluctuations with qubit performance parameters is key to achieving high-yield fabrication. This requires extensive manufacturing data accumulation and advanced statistical analysis methods~\cite{van2024advanced}.

\textbf{Ecosystem and standards gaps:} The standardization of data formats for quantum chip design, verification, and fabrication lags behind, hindering tool-chain interoperability and industry collaboration. For example, no unified exchange format for quantum chip PDKs currently exists, and data conversion between different EDA tools often requires custom adapter scripts.

Future directions include:

\textbf{AI-empowered intelligent MDP:} Leveraging machine learning to accelerate computationally intensive tasks such as OPC and hotspot detection, and to enable intelligent prediction and optimization of manufacturing yield. Deep learning models can learn process--performance mapping relationships from historical fabrication data, providing real-time feedback for design optimization.

\textbf{Cloud-native manufacturing file services:} Deploying the manufacturing file generation pipeline in the cloud to provide elastic computing resources and one-stop data delivery services, lowering the barrier to entry for small and medium-sized quantum computing teams.

\textbf{Design--fabrication collaboration platforms:} Building collaborative platforms connecting chip design companies, EDA tool vendors, and wafer foundries to enable real-time PDK updates, online design verification, and transparent fabrication status tracking.

\textbf{3D integration and multi-chip packaging:} As quantum processors evolve toward 3D integration (e.g., flip-chip~\cite{conner2021superconducting}, through-silicon vias~\cite{rosenberg20173d}) and multi-chip modules (MCMs)~\cite{das2024chip}, Q-EDA tools must support joint design of multi-layer chips, cross-chip signal integrity analysis, and 3D manufacturing file generation.

\section{Conclusion}

Wafer-scale fabrication is the inevitable path toward scaling superconducting quantum computing, and achieving accurate, efficient conversion from design layouts (GDSII) to executable manufacturing files is the core challenge that Q-EDA tools must overcome. This paper has systematically examined this data flow, covering PDK-based physical verification (DRC/LVS/ERC), multi-layer process stack mapping, DFM optimization, wafer layout planning, and mask data preparation.

Through the architecture of a self-developed Q-EDA system, we have presented nine quantum-specific DRC rules together with their physical underpinnings, and validated the completeness of the rule set through benchmarking against the CMC superconducting PDK white paper. Experimental results demonstrate that the Q-EDA data-flow processing achieves sub-millisecond response times and strict linear scalability, meeting the engineering requirements of large-scale wafer-level design.

Currently, tools such as Origin Pilot have made substantive progress along this path, but room for improvement remains in back-end stages such as DFM optimization and MDP. As the field moves toward the million-qubit era, advancing Q-EDA tools toward deeper integration with the fabrication end and establishing end-to-end ``design-to-fabrication'' data chains will be of significant importance for the quantum computing industry.

\bibliographystyle{IEEEtran}

\begin{thebibliography}{10}
\providecommand{\url}[1]{#1}
\csname url@samestyle\endcsname
\providecommand{\newblock}{\relax}
\providecommand{\bibinfo}[2]{#2}
\providecommand{\BIBentrySTDinterwordspacing}{\spaceskip=0pt\relax}
\providecommand{\BIBentryALTinterwordstretchfactor}{4}
\providecommand{\BIBentryALTinterwordspacing}{\spaceskip=\fontdimen2\font plus
\BIBentryALTinterwordstretchfactor\fontdimen3\font minus
  \fontdimen4\font\relax}
\providecommand{\BIBforeignlanguage}[2]{{%
\expandafter\ifx\csname l@#1\endcsname\relax
\typeout{** WARNING: IEEEtran.bst: No hyphenation pattern has been}%
\typeout{** loaded for the language `#1'. Using the pattern for}%
\typeout{** the default language instead.}%
\else
\language=\csname l@#1\endcsname
\fi
#2}}
\providecommand{\BIBdecl}{\relax}
\BIBdecl

\bibitem{zhao2025eda}
B.~Zhao, Z.~Li, X.~Yu, B.~Yuan, C.~Zhang, Y.~Gao, W.~Wang, Q.~Mu, S.~Wang,
  H.~Sun, T.~Yang, M.~Zhang, C.~Han, P.~Xu, W.~Wang, and Z.~Shan, ``{EDA-Q}:
  Electronic design automation for superconducting quantum chip,'' \emph{IEEE
  Transactions on Computer-Aided Design of Integrated Circuits and Systems},
  2025.

\bibitem{arute2019quantum}
F.~Arute, K.~Arya, R.~Babbush, D.~Bacon, J.~C. Bardin, R.~Barends, R.~Biswas,
  S.~Boixo, F.~G. Brandao, D.~A. Buell \emph{et~al.}, ``Quantum supremacy using
  a programmable superconducting processor,'' \emph{Nature}, vol. 574, no.
  7779, pp. 505--510, 2019.

\bibitem{van2024advanced}
J.~Van~Damme, S.~Massar, R.~Acharya, T.~Ivanov, D.~Perez~Lozano, Y.~Canvel,
  M.~Demarets, D.~Vangoidsenhoven, Y.~Hermans, J.~Lai \emph{et~al.}, ``Advanced
  cmos manufacturing of superconducting qubits on 300 mm wafers,''
  \emph{Nature}, vol. 634, no. 8032, pp. 74--79, 2024.

\bibitem{gdsii_format}
{Calma Company}, ``{GDSII} stream format manual,'' 1987, release 6.0.

\bibitem{gao2021practical}
Y.~Y. Gao, M.~A. Rol, S.~Touzard, and C.~Wang, ``Practical guide for building
  superconducting quantum devices,'' \emph{PRX Quantum}, vol.~2, no.~4, p.
  040202, 2021.

\bibitem{zheng2023fabrication}
Y.~Zheng \emph{et~al.}, ``Fabrication of al/alox/al junctions with high
  uniformity and reproducibility,'' \emph{Scientific Reports}, vol.~13, no.~1,
  p. 9636, 2023.

\bibitem{levenson2024review}
E.~M. Levenson-Falk and S.~A. Shanto, ``A review of design concerns in
  superconducting quantum circuits,'' \emph{Materials for Quantum Technology},
  2025, arXiv:2411.16967.

\bibitem{krylov2021review}
G.~Krylov, J.~Kawa, and E.~G. Friedman, ``Design automation of superconductive
  digital circuits: A review,'' \emph{IEEE Nanotechnology Magazine}, vol.~15,
  no.~6, pp. 54--67, 2021.

\bibitem{kjaergaard2020superconducting}
M.~Kjaergaard, M.~E. Schwartz, J.~Braum{\"u}ller, P.~Krantz, J.~I.-J. Wang,
  S.~Gustavsson, and W.~D. Oliver, ``Superconducting qubits: Current state of
  play,'' \emph{Annual Review of Condensed Matter Physics}, vol.~11, pp.
  369--395, 2020.

\bibitem{qiskitmetal2021}
{IBM Quantum}, ``Qiskit metal: An open-source framework for quantum device
  design \& analysis,'' \url{https://qiskit-community.github.io/qiskit-metal/},
  2021, accessed: 2026-03-01.

\bibitem{van2019layout}
R.~van Staden, J.~A. Delport, J.~A. Coetzee, and C.~J. Fourie, ``Layout versus
  schematic with design/magnetic rule checking for superconducting integrated
  circuit layouts,'' in \emph{2019 IEEE International Superconductive
  Electronics Conference (ISEC)}.\hskip 1em plus 0.5em minus 0.4em\relax IEEE,
  2019, pp. 1--5.

\bibitem{tolpygo2016advanced}
S.~K. Tolpygo, ``Advanced fabrication processes for superconducting very
  large-scale integrated circuits,'' \emph{IEEE Transactions on Applied
  Superconductivity}, vol.~26, no.~3, pp. 1--10, 2016.

\bibitem{gdsfactory2023}
{GDSFactory Contributors}, ``{GDSFactory}: {Python} library for chip design,''
  \url{https://gdsfactory.github.io/}, 2023, accessed: 2026-03-01.

\bibitem{kohli2023white}
N.~Kohli, C.~Paradis, S.~Grayli, and U.~Mendes, ``White paper: A process design
  kit for superconducting components,'' CMC Microsystems, Tech. Rep., 2023.

\bibitem{goppl2008coplanar}
M.~G{\"o}ppl, A.~Fragner, M.~Baur, R.~Bianchetti, S.~Filipp, J.~Fink, P.~Leek,
  G.~Puebla, L.~Steffen, and A.~Wallraff, ``Coplanar waveguide resonators for
  circuit quantum electrodynamics,'' \emph{Journal of Applied Physics}, vol.
  104, no.~11, p. 113904, 2008.

\bibitem{fourie2020eda}
C.~J. Fourie, ``Electronic design automation tools for superconducting
  circuits,'' \emph{Journal of Physics: Conference Series}, vol. 1590, no.~1,
  p. 012040, 2020.

\bibitem{huang2021microwave}
S.~Huang, B.~Lienhard, G.~Calusine, A.~Veps{\"a}l{\"a}inen, J.~Braum{\"u}ller,
  D.~K. Kim, A.~J. Melville, B.~M. Niedzielski, J.~L. Yoder, B.~Kannan
  \emph{et~al.}, ``Microwave package design for superconducting quantum
  processors,'' \emph{PRX Quantum}, vol.~2, no.~2, p. 020306, 2021.

\bibitem{place2021new}
A.~P. Place, L.~V. Rodgers, P.~Mundada, B.~M. Smitham, M.~Fitzpatrick, Z.~Leng,
  A.~Premkumar, J.~Bryon, A.~Vrajitoarea, S.~Sussman \emph{et~al.}, ``New
  material platform for superconducting transmon qubits with coherence times
  exceeding 0.3 milliseconds,'' \emph{Nature Communications}, vol.~12, no.~1,
  p. 1779, 2021.

\bibitem{chen2014fabrication}
Z.~Chen, A.~Megrant, J.~Kelly, R.~Barends, J.~Bochmann, Y.~Chen, B.~Chiaro,
  A.~Dunsworth, E.~Jeffrey, J.~Mutus \emph{et~al.}, ``Fabrication and
  characterization of aluminum airbridges for superconducting microwave
  circuits,'' \emph{Applied Physics Letters}, vol. 104, no.~5, p. 052602, 2014.

\bibitem{kqcircuits2022}
{IQM Finland}, ``{KQCircuits}: {KLayout} {Python} library for automating the
  design of superconducting quantum circuits,''
  \url{https://github.com/iqm-finland/KQCircuits}, 2022, accessed: 2026-03-01.

\bibitem{conner2021superconducting}
C.~R. Conner, A.~Bienfait, H.-S. Chang, M.-H. Chou, {\'E}.~Dumur, J.~Grebel,
  G.~A. Peairs, R.~G. Povey, H.~Yan, Y.~P. Zhong \emph{et~al.},
  ``Superconducting qubits in a flip-chip architecture,'' \emph{Applied Physics
  Letters}, vol. 118, no.~23, p. 232602, 2021.

\bibitem{rosenberg20173d}
D.~Rosenberg, D.~Kim, R.~Das, D.~Yost, S.~Gustavsson, D.~Hover, P.~Krantz,
  A.~Melville, L.~Racz, G.~Samach \emph{et~al.}, ``3d integrated
  superconducting qubits,'' \emph{npj Quantum Information}, vol.~3, no.~1,
  p.~42, 2017.

\bibitem{das2024chip}
A.~Das, M.~Palesi, J.~Kim, and P.~P. Pande, ``Chip and package-scale
  interconnects for general-purpose, domain-specific, and quantum computing
  systems---overview, challenges, and opportunities,'' \emph{IEEE Journal on
  Emerging and Selected Topics in Circuits and Systems}, vol.~14, no.~3, pp.
  354--370, 2024.

\end{thebibliography}

\end{document}